\documentclass[journal=jpclcd,manuscript=letter]{achemso} 

\usepackage[version=3]{mhchem} 
\usepackage{amsmath,amssymb}
\usepackage{xcolor} 

\newcommand{\affA}{Max Planck Institute for the Structure and Dynamics of Matter and Center for Free-Electron Laser Science, Luruper Chaussee 149, 22761 Hamburg, Germany
}
\newcommand{\affB}{Institute of Theoretical and Computational Chemistry, Universitat de Barcelona, Martí i Franquès 1, 08028 Barcelona, Spain.
}
\newcommand{\affC}{ICFO-Institut de Ciencies Fotoniques, The Barcelona Institute of Science and Technology, Castelldefels (Barcelona), E-08860, Spain}
\newcommand{\affD}{Departament de Ciència de Materials i Química Física, Universitat de Barcelona, Martí i Franquès 1, 08028 Barcelona, Spain.}
\newcommand{\affE}{Departament de Qu\'imica Inorg\`anica i Org\`anica, Universitat de Barcelona, Martí i Franquès 1, 08028 Barcelona, Spain.}
\newcommand{\affF}{IBM Research GmbH, Zurich Research Laboratory, Säumerstrasse 4, 8803 Rüschlikon, Switzerland
}

\author{Guillermo Albareda}\affiliation{\affA}\alsoaffiliation{\affB}
\email{guillermo.albareda@mpsd.mpg.de}
\author{Arnau Riera}\affiliation{\affC}
\author{Miguel Gonz\'alez}\affiliation{\affD}\alsoaffiliation{\affB}
\author{Josep Maria Bofill}\affiliation{\affE}\alsoaffiliation{\affB}
\author{Iberio de P. R Moreira}\affiliation{\affD}
\author{Rosendo Valero}\affiliation{\affD}
\author{Ivano Tavernelli}\affiliation{\affF}

\title[Quantum equilibration]
  {Quantum equilibration of a model system Porphine}

\abbreviations{IR,NMR,UV}
\keywords{American Chemical Society, \LaTeX}

\begin{document}

\begin{tocentry}
    \includegraphics[height = 5.1cm,width = 5.1cm]{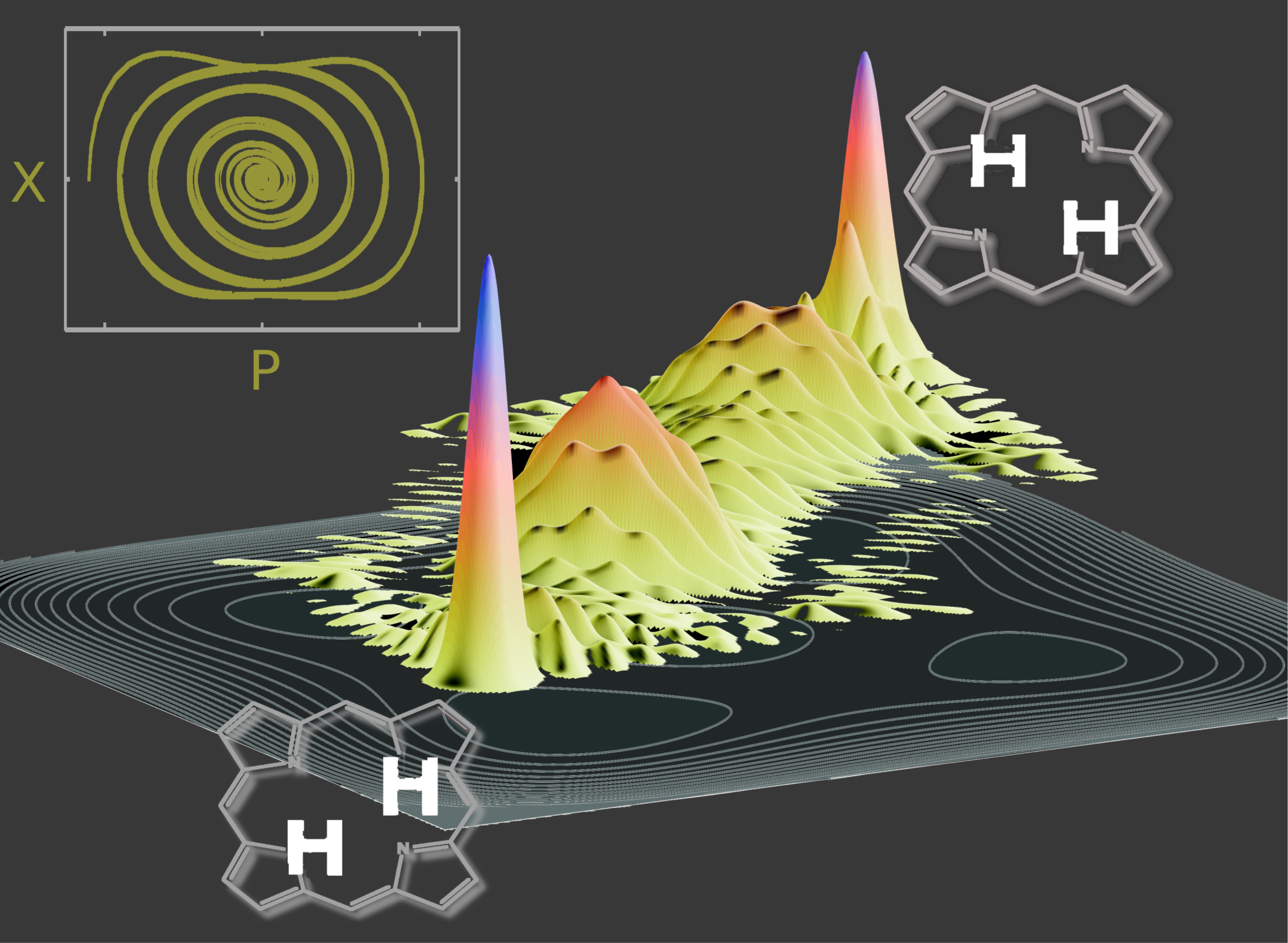}
\end{tocentry}

\begin{abstract}
There is a renewed interest in the derivation of statistical mechanics from the dynamics of closed quantum systems. 
A central part of this program is to understand how far-from-equilibrium closed quantum system can behave as if relaxing to a stable equilibrium.
Equilibration dynamics has been traditionally studied with a focus on the so-called quenches of large-scale many-body systems.
Alternatively, we consider here the equilibration of a molecular model system describing the double proton transfer reaction in porphine.
Using numerical simulations, we show that equilibration in this context indeed takes place and does so very rapidly ($\sim \!\! 200$fs) for initial states induced by pump-dump laser pulse control with energies well above the synchronous tunneling barrier. 
\end{abstract}

\textbf{Introduction:}
There is currently a renewed interest in the derivation of statistical mechanics from the dynamics of a closed quantum system~\cite{gogolin2016equilibration}. In this approach, instead of assuming a priori that the system is in some mixed state, such as e.g. a micro-canonical ensemble, one describes it at all times using a pure state. One then seeks to show that, under reasonable conditions, the system behaves as if it were described by a statistical ensemble. In this way the use of statistical mechanics can be justified without introducing additional external degrees of freedom, such as e.g. thermal ``baths''.

A central part of this programme has been to understand the process of equilibration, i.e., how a constantly-evolving closed quantum system can behave as if relaxing to a stable equilibrium. The main insight relies on the fact~\cite{reimann2007typicality,goldstein2006canonical,popescu2006entanglement} that, if measurements are limited to small subsystems or restricted sets of observables, then ``typical'' pure states of large quantum systems are essentially indistinguishable from thermal states. It can then be shown~\cite{reimann2008foundation,linden2009quantum} that under very general conditions on the Hamiltonian and nearly all initial states, the system will eventually equilibrate, in the sense that an (again, restricted) set of relevant physical quantities will remain most of the time very close to fixed, ``equilibrium'' values. 

Quantum equilibration has been recently put within reach of experimental verification. Fueled by enormous improvements in experimental techniques it is now feasible to control quantum systems with many degrees of freedom. This is particularly true for the development of techniques to cool and trap ultracold atoms and to subject them to optical lattices~\cite{greiner2002collapse,greiner2002quantum,tuchman2006nonequilibrium,bloch2005ultracold,langen2015ultracold}, giving rise to low-dimensional continuous systems~\cite{sadler2006spontaneous,kinoshita2006quantum,hofferberth2007non,weller2008experimental}. Similarly, systems of trapped ions~\cite{porras2004effective,friedenauer2008simulating} allow us to precisely study the physics of interacting systems in the laboratory~\cite{haffner2005scalable,jurcevic2014quasiparticle,lanyon2011universal,schindler2013quantum,blatt2012quantum,britton2012engineered}. In such highly controlled settings, equilibration and thermalisation dynamics has been studied both experimentally~\cite{trotzky2012probing,cheneau2012light,langen2013local,gring2012relaxation,ronzheimer2013expansion} and numerically, often with a focus on the so-called quenches, i.e. rapid changes of the Hamiltonian~\cite{rigol2008thermalization,rigol2009breakdown,moeckel2008interaction,kollath2007quench,deng2011dynamical,venuti2010universality,goth2012time,rigol2011initial,torres2014quench}. 

Here we aim at studying equilibration in a different context. 
We are interested in showing whether a small molecular system involving a few number of degrees of freedom might also equilibrate.
To answer the above question, we consider a model system describing the double proton transfer reaction in the electronic ground state of porphine, a paradigmatic system in which the making and breaking of H-bonds occurs in a highly anharmonic potential Born-Oppenheimer energy surface. 
Despite the small number of degrees of freedom involved (only two protons for this particular model system), we will see that, for initial states that involve a large coherent sum of vibrational eigenstates, the isomerization dynamics can reach a long-lived quasi-stationary regime where physical quantities such as the mean position or momentum remain very close to fixed, ``equilibrium'', values.

\textbf{Definitions:}
The notion of equilibration is compatible with the recurrent and time reversal invariant nature of unitary quantum dynamics in finite dimensional systems. This notion captures the intuition that equilibration means that a measurable quantity, after having been initialised at a non-equilibrium value, evolves towards some value and then stays close to it for an extended amount of time. 

Given some observable $A$ and a system of finite but arbitrarily large size, if its expectation value $\langle \hat A(t)\rangle $ equilibrates, then it must do so around the infinite time average~[1],
\begin{equation}
\bar{A} = \lim_{T\to\infty} \frac{1}{T}\int_0^T \langle \psi(t) | \hat A | \psi(t) \rangle dt.
\end{equation}
If the infinite-time average fluctuation of $\langle \hat A(t)\rangle $ around $\bar{A}$ is small, then we say that the observable $A$ equilibrates.

For a closed system whose state is described by a vector in a Hilbert space of dimension $d_T$ and whose Hamiltonian has a spectral representation
$\hat H = \sum_{k=1}^{d_T} E_k |E_k\rangle \langle E_k|$,
where $E_k$ are its energies and $|E_k\rangle$ are eigentstes of $\hat H$. \footnote{Note that the sum runs over $d_E \leq d_T$ terms, since some eigenspaces can be degenerate.
If the Hamiltonian has degenerate energies, we choose an eigenbasis of $\hat H$ such that the initial state, $| \psi (0)\rangle $, has non-zero overlap with only one eigenstate $| {E}_{k}\rangle $ for each distinct energy.} 
Choosing units such that $\hbar = 1$, the state at time $t$ is then given by
$|\psi(t)\rangle = \sum_k c_k e^{iE_k t} |E_k\rangle$
with ${c}_{k}\equiv \langle {E}_{k}| \psi (0)\rangle $. Therefore, if the system equilibrates, the equilibrium state must be described by the following density matrix:
\begin{equation}
 \hat \omega = \sum_k |c_k|^2 |E_k\rangle \langle E_k|.
\end{equation}
Note that what we call equilibration is less than what one usually associates with the evolution towards thermal equilibrium.

Sufficient conditions for equilibration in the sense defined above can be then defined in terms of the so-called effective dimension. Roughly, the effective dimension is a measure of the number of signinficantly occupied energy eigenstates, and is defined as:
\begin{equation}\label{deff}
    d_{\rm{eff}}^{-1}:= \sum_k |c_k|^4 = \text{Tr}\left(\omega^2\right)
\end{equation}
In Ref.~\cite{reimann2008foundation,linden2009quantum,Short_2011} it is shown that for any Hamiltonian with non-degenerate gaps\footnote{Concerning the assumption of the Hamiltonian having non-degenerate gaps, it is shown in \cite{Short_2012} that as long as there are not exponentially many degeneracies the argument stays the same.},
a large effective dimension is sufficient to guarantee that equilibration will be attained.
Note that if the initial state is taken to be an energy eigenstate, the resulting effective dimension is one, while the one resulting from a uniform coherent superposition of $d$ energy eigenstates to different energies is $d_T$. 
There are a number of different ways to argue why it is acceptable to restrict oneself to initial states that populate a large number of energy levels when trying to prove the emergence of thermodynamic behaviour from the unitary dynamics of closed systems~\cite{gogolin2016equilibration}.

Provided that a given observable equilibrates, it was recently shown~\cite{de_Oliveira_2018} that the equilibration time can be related with a dephasing time $\tau_d$ as:
\begin{equation}\label{teq}
    T_{eq} \sim \tau_d = \pi/\sigma_G.
\end{equation}
where $\sigma_G$ is the standard deviation of the energy gaps, $G_\alpha = E_j-E_i$, when weighted by their respective relevances $q_\alpha$, i.e.:
\begin{equation}
    \sigma_G^2 := \sum_\alpha q_\alpha G_\alpha^2,
\end{equation}
where $\alpha \in { \mathcal G }=\{(i,j):i,j\in \{1,\,\ldots \,{d}_{E}\},i\ne j\}$, and $q_\alpha := |v_\alpha|^2/\sum_\beta |v_\beta|^2$ with $v_\alpha = {c_j^*A_{ji}c_i}/{\Delta_A}$, and $A_{ij}:=\langle E_i|A|E_j \rangle$ are the matrix elements of $A$ in the energy eigenbasis. The denominator $\Delta_A = a_{max} - a_{min}$ is the range of possible outcomes, being ${a}_{\max (\min )}$ the largest (smallest) (occupied) eigenvalue of $\hat A$.

\begin{figure}
 \includegraphics[width=\textwidth]{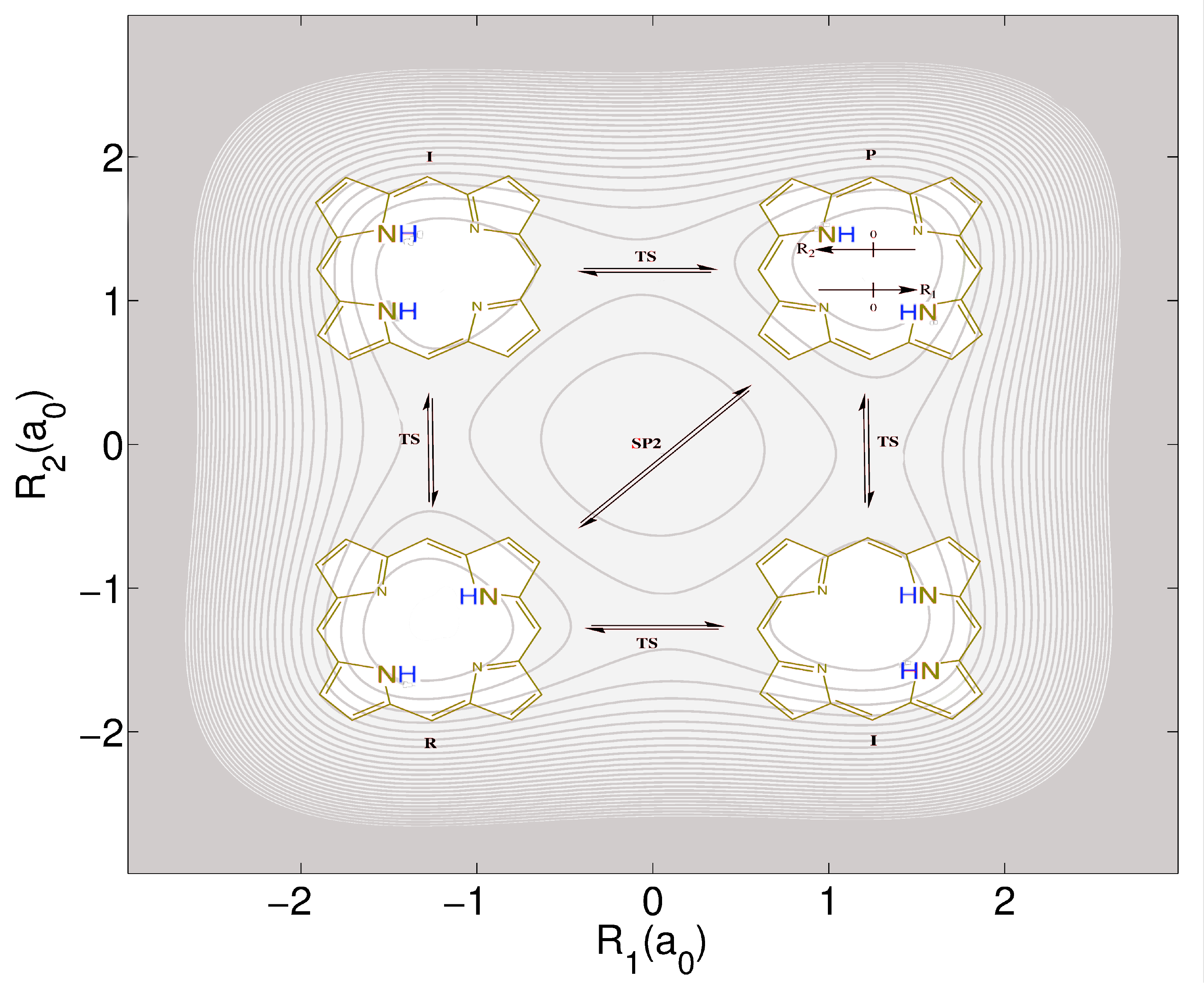}
 \caption{Double proton tranfer of the model porphine. The protons
move along coordinates $R_1$ and $R_2$. The four snapshots represent the
transfer of the two protons from reactant (R) to product (P),
sequentially along intermediate states (I) involving four transition
states (TS), or simultaneously through a second order saddle point
(SP2). In the background: Potential energy surface for the model porphine, Eq. \eqref{PES}, adopted from Refs.~\citenum{Smedarchina,Accardi,jpcl_guille}. The equidistant values of the contours range from 0eV, for the potential minima for the reactant (R) and product (P) configurations, to 5eV. The corresponding energies of the local minima for the intermediates (I), of the four barriers labeled TS, and of the second order saddle point (SP2) are 0.238, 0.600, and 1.069eV, respectively.}
 \label{porphine_scheme}
\end{figure}

\textbf{The model Porphine:}
The above introduced concepts are now used to characterize the model Porphine designed by \citet{Smedarchina} to describe the switch from synchronous (or concerted) to sequential (or stepwise) double-proton transfer\cite{Accardi,jpcl_guille}. This model accounts for the motions of two protons (labeled $1$ and $2$) along
coordinates $R_1$ and $R_2$, respectively, from the domains of the reactant (R) to the product (P) (see Fig.\ref{porphine_scheme}).
The PES model is\cite{Smedarchina}  
\begin{equation}\label{PES}
 V(R_1,R_2) = \frac{U_0}{\Delta_0^4}\Big[  (R_1^2 - \Delta_0^2)^2  +  (R_2^2 - \Delta_0^2)^2  - 4G\Delta_0^2R_1R_2  \Big]  +  2G(2+G)U_0.
\end{equation}
The parameter $U_0 = 0.473$eV has been fitted in Ref.\citenum{Smedarchina} in order to account for the experimental results of nuclear magnetic resonance and laser-induced fluorescence measurements
of Refs.\citenum{Butenhoff,Braun1,Braun2}. The other two parameters, $\Delta_0 = 1.251a_0$ and $G = 0.063$ are based on density functional theory calculations of \citet{Smedarchina2} at 
the B3LYP/6-31G* level. 
The resulting 2D model PES is illustrated in the background of Figure \ref{porphine_scheme}.
The barriers are labeled TS (``transition states'') for two alternative reaction paths. 
The reaction can lead from the reactant R via alternative transitions states TS to the intermediates (I), and subsequently via the other two TS to the product P.
In addition, Fig.\ref{porphine_scheme} shows a central saddle point (of second order) labeled SP2. The competing
synchronous reaction mechanism leads from the reactant R via SP2 to the product P.

The model potential in \eqref{PES} is symmetric with respect to the diagonals $R_1 = \pm R_2$. It accommodates
nearly degenerate doublets of eigenstates $\Psi_{v+}(R_1, R_2)$ and $\Psi_{v-}(R_1, R_2)$, with energies below the barriers TS,
plus higher excited states. 
We then chose our initial state to be of the general form:
\begin{equation}\label{initial_state}
    \Psi_0(R_1,R_2;\Delta R) = \Psi_{0,R}(R_1 + \Delta R,R_2 + \Delta R),
\end{equation}
where $\Psi_{0,R}(R_1,R_2) = \frac{1}{\sqrt{2}}(\Psi_{0+} + \Psi_{0-})$
is a superposition state that represents the localized ground state wave function of the reactant, where $\Psi_{0+}(R_1, R_2)$ and $\Psi_{0-}(R_1, R_2)$ represent the lowest doublet ($v = 0$).
This type of shifts of initial wavefunctions from equilibrium to non-equilibrium positions may be induced, for example, by means of pump-dump
laser pulse control, as designed by Tannor and Rice~\cite{tannor1985control,tannor1986coherent,rice2000optical}. Essentially, the ultrashort pump pulse transfers the molecule from the electronic ground state to an excited state. Here, the system evolves from the original configuration until it is shifted to the target position. Finally, the dump pulse sets the wavepacket back to the electronic ground state, thus preparing the initial state for the subsequent reaction. Analogous shifts of the original wavefunctions to nonequilibrium positions have been demonstrated recently by means of laser pulse control, by Kapteyn, Murnane, and co-workers~\cite{li2008time}.

\begin{figure}
 \includegraphics[width=\textwidth]{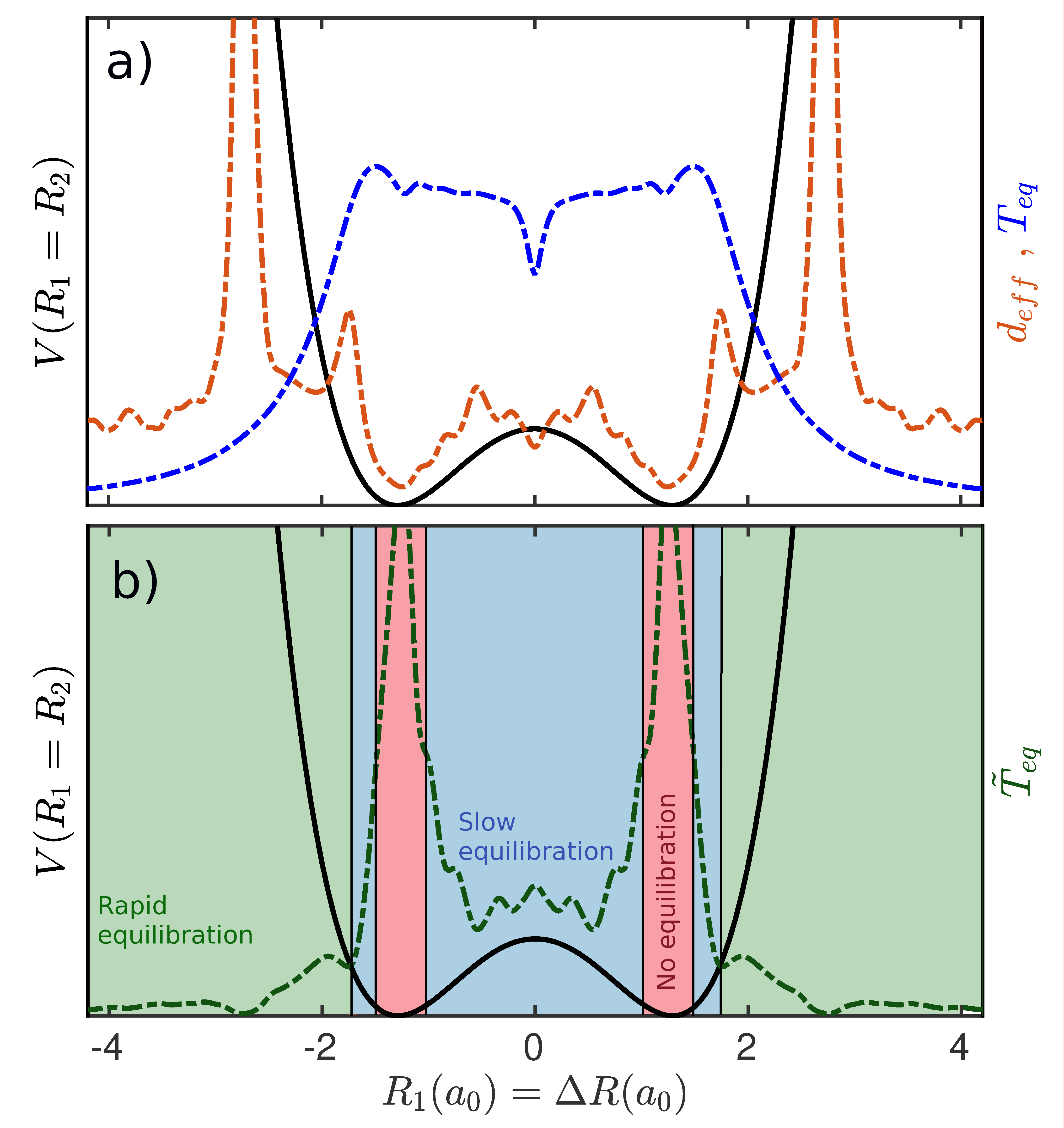}
 \caption{Panel a: Effective dimension $d_{eff}$ (in dashed red line) and equilibration time $T_{eq}$ (in dahsed blue line) for the initial states defined in Eq.~\eqref{initial_state} and different values of $\Delta R$. The equilibration time has been computed for the specific type of measurement defined in Eq.~\eqref{position_measure}.
 Panel b: The resulting effective equilibration time, $\tilde{T}_{eq}$, defined in Eq.~\eqref{teqeff}. Three different regions are defined: i) in green, regions where equilibration is expected to occur rapidly, ii) in blue, regions where equilibration it is expected to be attained slowly, and iii) in red, regions where equilibration is not expected to occur.}
 \label{equilibration_scheme}
\end{figure}

\textbf{Equilibration dynamics:}
We consider the type of projective position measurements
\begin{equation}\label{position_measure}
  \hat R = |R_1\rangle\langle R_1| \otimes \mathbb{I}_2 + \mathbb{I}_1 \otimes | R_2\rangle\langle R_2|,    
\end{equation}
where $\mathbb{I}_{1,2} = \sum_{-\infty}^\infty |R_{1,2}\rangle\langle R_{1,2}|$ are identity operators in the position representation acting, respectively, in the subspaces defined by $R_1$ and $R_2$. 
We then evaluate the effective dimension $d_{eff}$ as well as the equilibration time $T_{eq}$ in Eqs. \eqref{deff} and \eqref{teq} respectively for different values of $\Delta R \in [-6,6]$ (see the top panel of Fig.~\ref{equilibration_scheme}). 
For that, we first computed $v_\alpha = {c_j^*R_{ji}c_i}/{\Delta_R}$, where $R_{ij}:=\langle E_i|\hat R|E_j \rangle$ and $\Delta_R = R_{max} - R_{min}$ has been chosen to be $\Delta R = 3\sigma_{gs}$, and $\sigma_{gs}= 1/\sqrt{25}$ is the dispersion of the localized ground state wave function of the reactant. 

Defining an equilibration time does only make sense if equilibration occurs, so we found convenient at this point to define an effective equilibration time as 
\begin{equation}\label{teqeff}
    \tilde{T}_{eq} = T_{eq}/d_{eff},
\end{equation}
which takes into account whether or not the system is expected to equilibrate. Note that, by definition, even if $T_{eq}$ is small, $\tilde{T}_{eq}$ can still be close to infinite for very small $d_{eff}$.
The resulting effective equilibration time $\tilde{T}_{eq}$ is plot in the bottom panel of Fig.\ref{equilibration_scheme}. Three different regions can be identified: i) in green, regions where equilibration is expected to occur fast, ii) in blue, regions where equilibration is expected to occur slow, and iii) in red, regions where equilibration is not expected to occur. Interestingly, region (iii) corresponds to the regime where the Born-Oppenheimer potential energy surface can be well approximated by a quadratic potential and thus $d_{eff}$ takes very small values. Furthermore, as there is no dephasing mechanism for a harmonic oscillator, the equilibration time $T_{eq}$ is expected to be large. Therefore, the effective temperature $\tilde{T}_{eff}$ grows ``exponentially'' as $\Delta R \to 0$. 
As one departs from the harmonic region, $\tilde{T}_{eff}$ decreases rapidly (although not monotonically). Regions of slow equilibration (in blue) are the result of a non-trivial interplay between $T_{eff}$ and $d_{eff}$.
In the regions where equilibration is expected to occur fast (in green), the effective dimension $d_{eff}$ and time $T_{eff}$ increase and decrease respectively, and thus $\tilde{T}_{eff}$ decreases too. Note that while the equilibration time decreases monotonically for $|\Delta R| \gtrsim 1.6$, the effective dimension reaches a maximum at $|\Delta R| \sim 2.6$ and decreases afterwards. This can be explained by the fact that at very high energies the double well potential can be well approximated by a single well potential (i.e., a harmonic oscillator with a natural frequency that is approximately half the frequency of each of the double well).   


In what follows we aim at studying the dynamics that leads to equilibration for initial states 
induced by pump-dump laser pulse control, i.e., of the form \eqref{initial_state}. 
Specifically, we choose $\Delta R = -2.2a_0$ such that the initial state lies in the rapid equilibration regime (see Fig.\ref{equilibration_scheme}). The mean energy of this initial state is $4.885$eV, i.e., well above the values of the barriers TS ($0.600$eV), and also the saddle point SP2, ($1.069$eV). 

\begin{figure}
 \includegraphics[width=\textwidth]{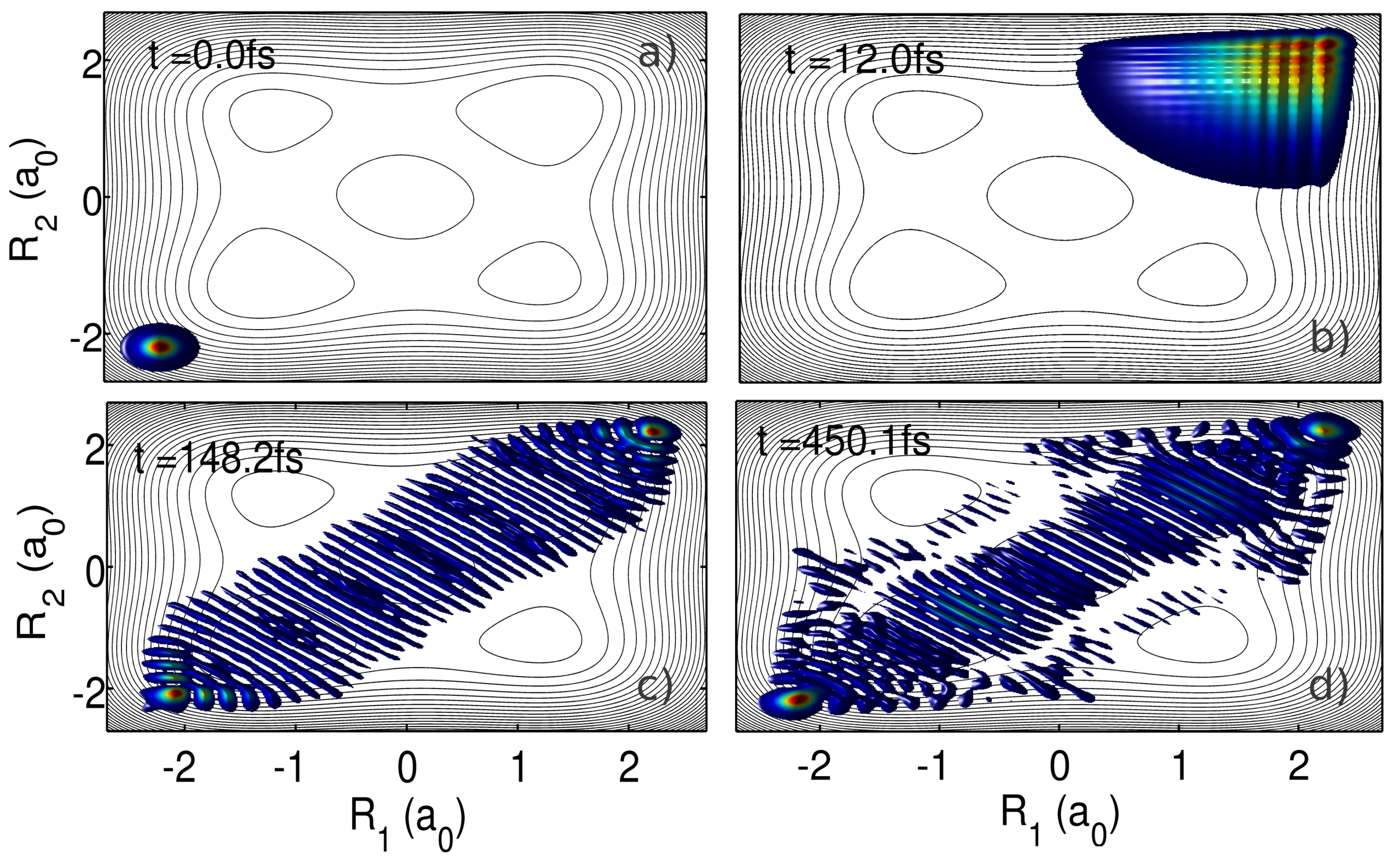}
 \caption{Dynamics of the two-proton density $|\Psi(R_1,R_2,t)|^2$ at four different time snapshots. The initially well-localized nuclear density gives way later to a strong proton delocalization along the synchronous pathway that is maintained for at least 1ps.}
 \label{dens_dynamics}
\end{figure}

Starting with $\Psi_0(R_1,R_2;-2.2a_0)$ (see Fig.~\ref{dens_dynamics}.a), the initial synchronous mechanism of the first forward reaction is characterized by a rapid dispersion of the wavepacket and relief reflections of the broadened wavepacket from wide regions of the steep repulsive wall of the PES close to the minimum of the product that leads to the switch from the synchronous to sequential mechanism (see Fig. \ref{dens_dynamics}.b). This situation is reminiscent of the near field interference effect arising when periodic diffracting structures are
illuminated by highly coherent light or particle beams\cite{talbot}.  
The time dilatation supported by continuous wavepacket dispersion leads to a strong proton delocalization (at times $t\gtrsim 100$fs).
An apparently ``chaotic'' flux is however fully coherent and ultimately directs the recovery of the concerted double proton transfer at $t\gtrsim 100$fs. Due to the strong time dilatation between partial waves, the grid structure of the probability density associated to the sequential double proton transfer progressively dilutes into what reminds a stationary state, showing a series of minima disposed perpendicular to the diagonal $R_1 = R_2$ (see Fig.~\ref{dens_dynamics}.c). 

\begin{figure}
 \includegraphics[width=\textwidth]{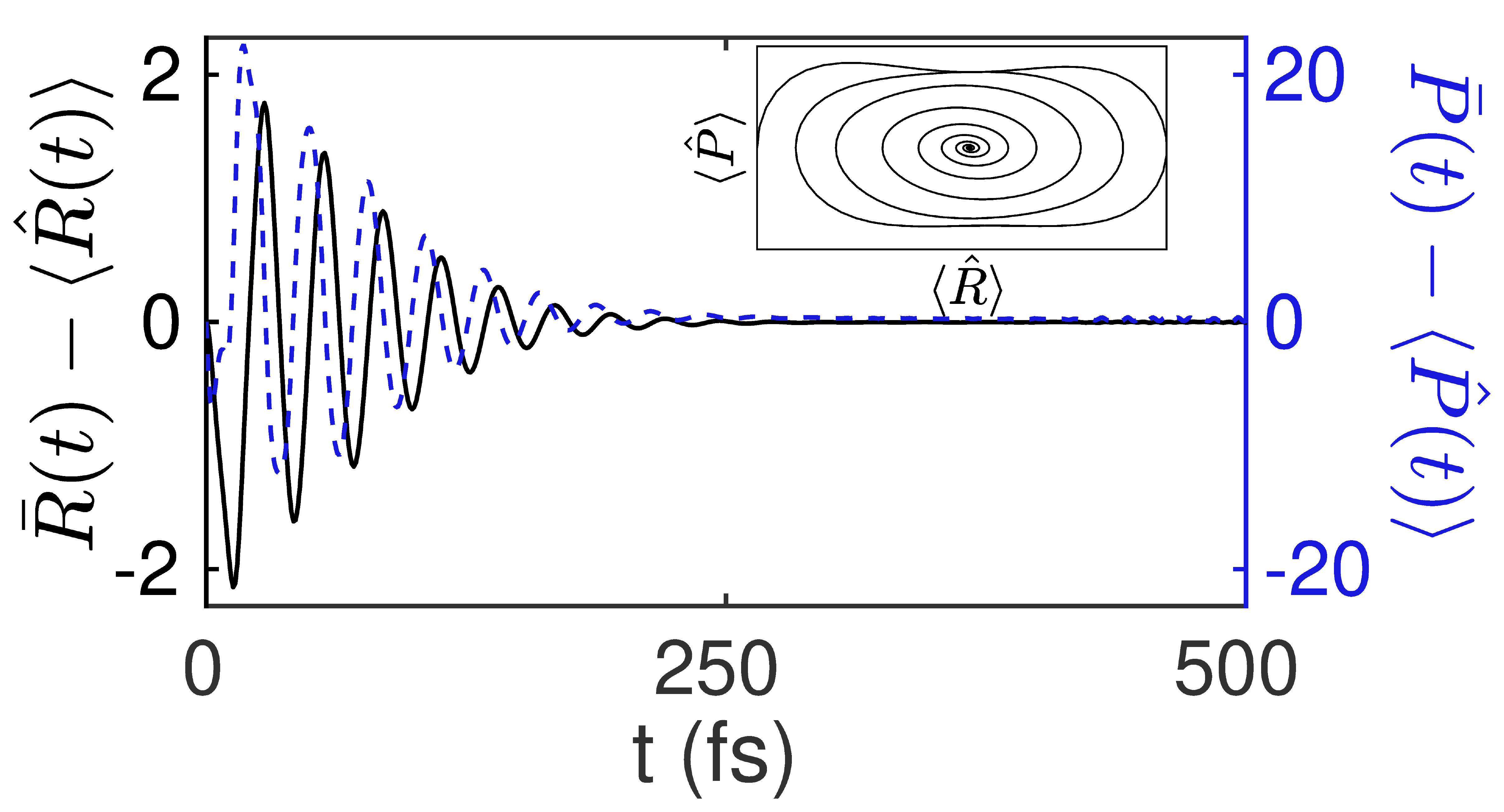}
 \caption{Equilibration of the position (solid black line) and momentum (dashed blue line). In the inset: ensemble average of the phase space as a function of time.}
 \label{observables}
\end{figure}

The long-lived quasi-stationary state, formed already at $~150$fs, lasts beyond the $~500$fs (see Fig.~\ref{dens_dynamics}.d) and it gives raise to the equilibration of the position operator in \eqref{position_measure} as well as of the momentum operator $\hat P = |P_1\rangle\langle P_1| \otimes \mathbb{I}_2 + \mathbb{I}_1 \otimes | P_2\rangle\langle P_2|$. In Figure ~\ref{observables}) we show the differences $\bar R(t) - \langle \hat R(t) \rangle$ and $\bar P(t) - \langle \hat P(t) \rangle$ as a function of time. Equilibration is achieved once these differences become small ($\ll 1$) and stay small for a lapse of time that is comparable to the relevant time-scale of the dynamics~\cite{malabarba2015rapid}. 
The equilibration of the position and momentum yields a phase-space diagram (see the inset in Fig.~\ref{observables}) that is reminiscent of the phase-space portrait of a damped harmonic oscillator. Interestingly, this phase-space dynamics is attained here without any loss (nor gain) of energy. 

To gain more insight into the physical mechanism that yields the phase-space portrait in Fig.~\ref{observables}, we here take advantage of the interpretation value of Bohmian mechanics~\cite{ApplBohm,BookOriols}. We introduce an ensemble of Bohmian trajectories $\{ \mathbf{R}^{\alpha}(t)\} =  \{ R_1^{\alpha}(t),R_2^{\alpha}(t)\}$ initially sampled from $|\Psi_0(R_1,R_2)|^2$ and whose evolution is defined as: 
  $\mathbf{R}^\alpha(t) = \mathbf{R}^\alpha(t_0) + \int_{t_0}^t \mathbf{v} (\mathbf{R}^\alpha(t'),t') dt'$,
where $\mathbf{v} (\mathbf{R}_1,\mathbf{R}_2,t) = (\nabla_{R_1} S/{M},\nabla_{R_2} S/{M})$
is the Bohmian velocity field, and $S(\mathbf{R}_1,\mathbf{R}_2,t)$ is the phase of the wavefunction $\Psi(R_1,R_2,t)$. 
\begin{figure}
 \includegraphics[width=\textwidth]{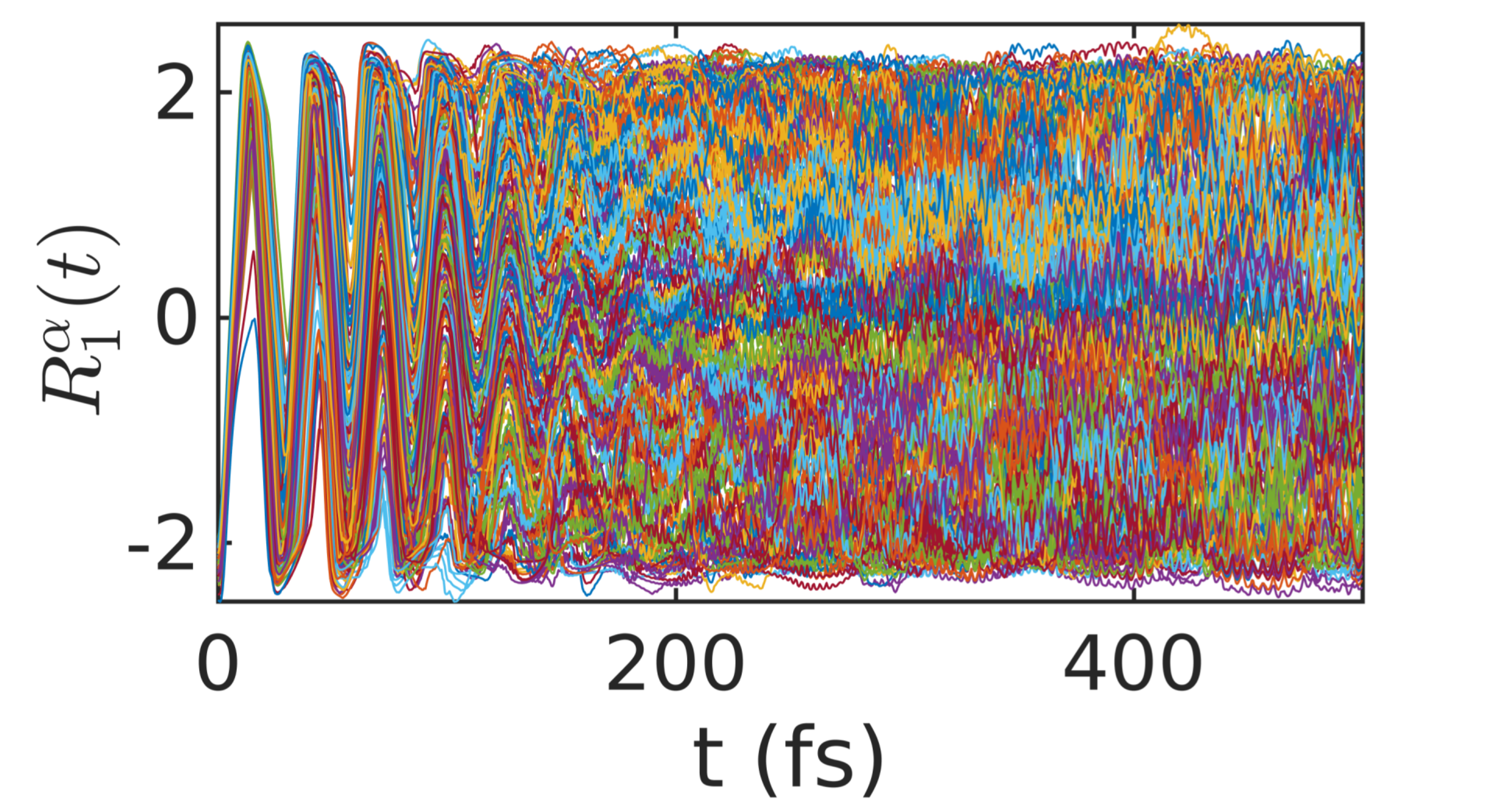}
 \includegraphics[width=\textwidth]{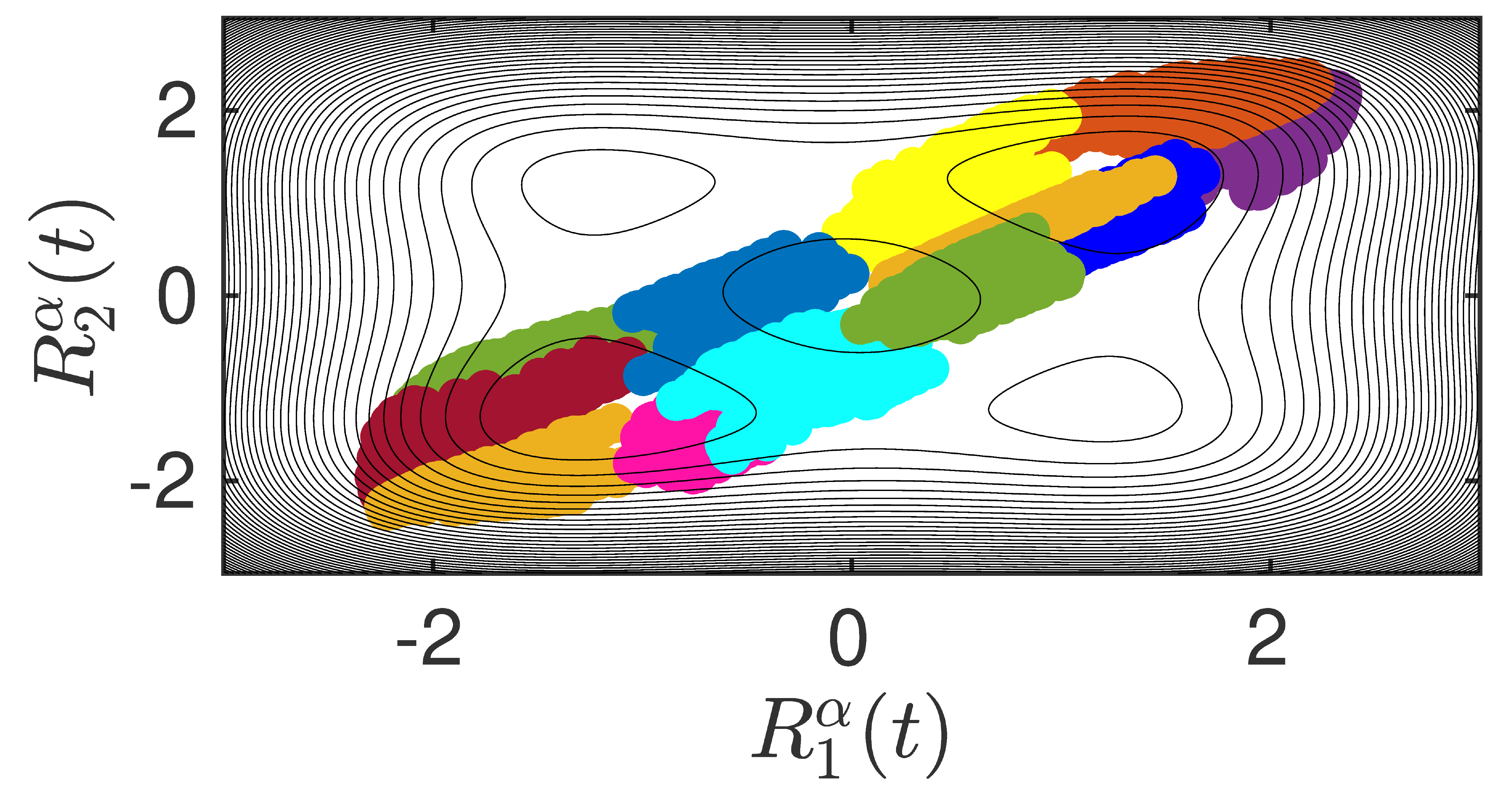}
 \caption{Top panel: time evolution of a randomly chosen sample of one thousand Bohmian trajectories $R_1^\alpha(t)$. Bottom panel: area explored by the motion of twelve randomly selected Bohmian trajectories $\mathbf{R}^\alpha(t)$ in the time interval $t\in[150, 500]$fs. A contour plot of the underlying potential energy surface has been added for the sake of clarity.}
 \label{trajectories}
\end{figure}
In the top panel of Fig.~\ref{trajectories} we show the time evolution of a (randomly chosen) sample of one thousand Bohmian trajectories $R_1^\alpha(t)$. An initially collective oscillatory dynamics with an amplitude that extends over the two wells is rapidly (at around $~$50fs) substituted by a more restricted oscillation within the wells. Not much later, at around 150fs, the trajectories fall into a pretty localized dynamics. 
Of particular interest is the fine structure of the Bohmian trajectories, characterized by high frequency components, and the sharp shifts in their direction. This quantum mechanical motion arises from strong time dependent quantum force fields which essentially dominates the dynamics and is responsible for shifting the direction of trajectories away from that dictated by
the classical force field. 

The above dynamics is the result of a very efficient dephasing mechanism. The barrier in between the two wells induces a continuous time dilatation supported by wavepacket dispersion that yields strong interference effects between different parts of the wavepacket. Eventually, these interferences are responsible for frizzing the transfer of the two protons and lead to the localization of the corresponding trajectories within the concerted pathway (see the bottom panel of Fig.~\ref{trajectories}). 
This equilibration dynamics corresponds to the picture where the two Hydrogen atoms become partially frozen at certain configurations along the concerted transfer path. 
Note that the close connection of Bohmian mechanics to weak values~\cite{wiseman2007grounding} provides an operational interpretation of the above trajectories, and puts them within reach of experimental verification~\cite{kocsis2011observing,mahler2016experimental}. 
In this respect, aside from its theoretical (and interpretative) value, Bohmian trajectories could offer new ways to experimentally quantify the role of equilibration.

\textbf{Conclusions:} To summarize, we have shown that quantum equilibration, mostly studied for many-body systems made of a large number of particles, can also play a role in molecular processes that involve a few number of degrees of freedom. 
Specifically, we have seen that a two-dimensional model system describing the double proton transfer in porphine can present equilibration dynamics for a wide range of far-from-equilibrium initial states that are compatible with a pump-dump preparation scheme. 
By picking up a specific initial condition that is expected to yield the rapid equilibration of the two protons, we have seen that the resulting equilibration state is reached in the femptosecond time scale and that it is associated with zero mean position and momentum of the two hydrogen atoms. 
Furthermore, this equilibration state is characterized by a strong delocalization of the probability density of the hydrogen atoms along the concerted transfer path.  
Relying on a Bohmian mechanics interpretation of the equilibration dynamics, we have shown that equilibration in porphine can be associated with the picture where the two Hydrogen atoms become ``frozen'' at certain configurations along the concerted transfer pathway.

\begin{acknowledgement}

G.A. acknowledges financial support from the European Unions Horizon 2020 research and innovation programme under the
Marie Skodowska-Curie Grant Agreement No. 752822, the
Spanish Ministerio de Economa y Competitividad (Project
No. CTQ2016-76423-P), and the Generalitat de Catalunya
(Project No. 2017 SGR 348).

\end{acknowledgement}


\bibliography{bibliography_equilibration.bib}

\end{document}